\documentclass[doublecol,linenumbers]{epl2} 

\usepackage{amssymb}
\usepackage{graphicx,xcolor}
\usepackage{epigraph}
\usepackage{csquotes}
\usepackage{amsmath,amssymb,stmaryrd}
\usepackage{amsmath,graphicx}
\def\la{{\langle}}
\def\u{\hat U}

\def\x{\overline}

\def\T{\mathcal{T} }

\def\D{\overline{D}}
\def\ot{\otimes}

\def\Om{ \Omega}
\def\om{ \omega}

\def\R{\text {Re}}
\def\Ip{\text {Im}}
\def\F{\overline F}
\def\W{\overline W}

\def\e{\enquote}

\def\q{\quad}

\def\n{\\ \nonumber}
\def\ra{{\rangle}}

\title{ Wigner's friends, tunnelling times and Feynman's \e{only mystery of quantum mechanics} }
\author{D. Sokolovski\inst{1,2}, and E. Akhmatskaya\inst{2,3} }
\institute{                    
\inst{1} IKERBASQUE, Basque Foundation for Science, E-48009 Bilbao, Bizkaia, Spain\\
   \inst{2} Departmento de Qu\'imica-F\'isica, Universidad del Pa\' is Vasco, UPV/EHU, Leioa, Bizkaia, Spain\\
\inst{3} Basque Center for Applied Mathematics (BCAM), Alameda de Mazarredo,
14, 48009 Bilbao, Bizkaia, Spain}
\abstract{
Recent developments in elementary quantum mechanics have seen a number of extraordinary claims 
regarding quantum behaviour, and even questioning internal consistency of the theory.
These are, we argue, different disguises of what Feynman described as quantum theory's \e{only mystery} }
\begin{document}
\maketitle
 \date\today

%
%
%
%
\epigraph{
Real mystics don't hide mysteries, they reveal them. They set a thing in broad daylight, and when you've seen it it's still a mystery. But the mystagogues hide a thing in darkness and secrecy, and when you find it, it's a platitude.}
{G. K. Chesterton in  {\it The arrow of heaven}}
\epigraph{I will take just one this experiment, which has been designed to contain all of the mysteries of quantum mechanics....
Any other situation in quantum mechanics, it turns out, can always be explained by saying \e{You remember the case of the experiment with the two holes?}}
{R.P. Feynman in {\it Character of Physical Law}}

\section{Introduction} We would like to begin by assuring the reader that the first item in the epigraph was not chosen in order to be offensive. 
However, the two quotes suit the purpose of this article so well that it would be a pity not to include them.
Since Wigner's surprising suggestion \cite{Wig} that the laws of quantum mechanics may need to be modified to accommodate human consciousness, 
there have been many other  eye-catching claims, associated with quantum  behaviour. The list includes, to name a few, particles being in several places 
at the same time \cite{3box}, electrons \e{disembodied of their charge} \cite{cat1}, quantum \e{Cheshire cats} \cite{cat}, photons with \e{discontinuous trajectories} \cite{phot}, observer-dependent facts \cite{bruckner},\cite{wiseman}, doubts about the internal consistency of elementary quantum theory \cite{renner} and the conflict of \e{faster-than-light tunnelling} with special relativity \cite{Nimtz}. A recent claim that it takes a finite amount of time to tunnel across a potential barrier 
can be found in \cite{Stein}.
\newline
Feynman, for his part, maintained that all rarities of quantum mechanics can be traced back to the double slit experiment.
So are these quantum \e{paradoxes} just different form of the familiar {double-slit conundrum?}
We show that most, if not all of them, are.
\section{The double-slit experiment. Feynman's rules}
It is worth recalling the description, given by Feynman in his undergraduate text \cite{FeynL}.
An electron, emitted by a source (s), can reach a point on a screen, $x$, by passing via two slits, labelled $1$ and $2$. 
If it is impossible {\it in principle} to establish which of the two paths was taken, the probability of arriving at $x$ is given by the absolute 
square of the sum of the path amplitudes, $P(x)=|A(x\gets 1\gets s)+A(x\gets 2\gets s)|^2$.
If it is possible, even {\it in principle}, the path probabilities are added, $P(x)=|A(x\gets 1\gets s)|^2+|A(x\gets 2\gets s)|^2$.
The fact that these two situations are mutually  exclusive constitutes the Uncertainty Principle (UP).

A useful illustration can be  provided by considering a two-level system (S) (a spin-$1/2$) and two probes $\x D$ and $D$, used to measure the spin's condition
at $t_1$ and $t_2 > t_1$, respectively. The joint initial state at some $t_0< t_1$ is $|\Psi(0)\ra=|D(0)\ra\ot|\x D(0)\ra\ot|s_0\ra$, and 
just after $t_1$ we have
\begin{eqnarray} \label{01}
|\Psi(t_1)\ra=|D(0)\ra\ot[\la up |s_0\ra |\x D(up)\ra\ot|up\ra +\n
\la down |s_0\ra|\x D(down)\ra\ot|down\ra],
\end{eqnarray}
where $|up\ra$ and $|down\ra$ form a measurement basis in the spin's Hilbert space, and we assumed, for simplicity, only the spin has its own dynamics. 
\begin{figure}
\onefigure[angle=0,width=6.0 cm, height= 3cm]{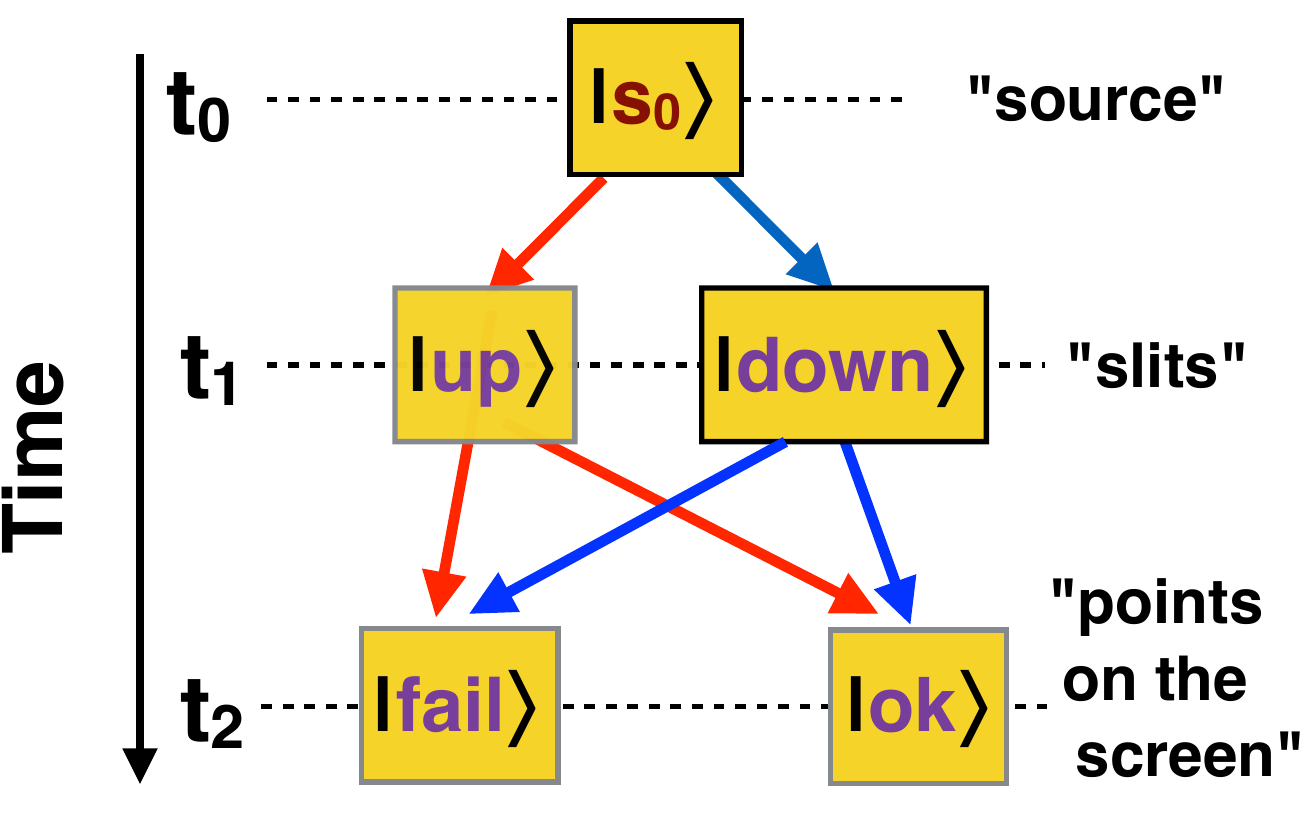}
\caption{A primitive double slit experiment. A spin can reach 
its final states via four virtual paths.  }
\label{fig.1}
\end{figure}
\newline
If the second measurement  engages only the spin, by using a measurement  basis,
\begin{eqnarray} \label{02}
|ok\ra =\alpha |up\ra+\beta|down\ra, \q |fail\ra =\gamma |up\ra+\delta|down\ra
\end{eqnarray}
the probability for each of the four possible outcomes is easily found to be 
\begin{eqnarray} \label{03}
P(i,j)=
|A^S(j\gets i\gets s_0)|^2,\n i=up,down, \q j=ok,fail,
\end{eqnarray}
where 
\begin{eqnarray} \label{04}
A^S(j\gets i\gets s_0)\equiv \la j | \u^S(t_2,t_1)|i\ra\la i | \u^S(t_1,t_0)|s_0\ra,
\end{eqnarray}
and $\u^S(t',t)$ is the spin's evolution operator.
If the second measurement engages both the spin and the probe $\D$, 
by using a basis containing the states
\begin{eqnarray} \label{05}
|ok\ra=\alpha|\x D(up)\ra\ot|up\ra + 
\beta |\x D(down)\ra\ot|down\ra,\n
|fail\ra=\gamma |\x D(up)\ra\ot|up\ra + 
\delta |\x D(down)\ra\ot|down\ra,
\end{eqnarray}
the probabilities for the two corresponding outcomes are given by 
\begin{eqnarray} \label{08}
P(j)=
|A^{S}(j \gets up\gets  s_0)+
A^{S}(j \gets down\gets  s_0)|^2,\n
j=ok,fail.
\end{eqnarray}
In the first case, system's past can be determined by inspecting the state of the probe $\D$ after the experiment is finished.
Accordingly, the four scenarios in Fig. 1 can be distinguished, and endowed with the probabilities in Eq.(\ref{03}). 
In the second case, the record carried by $\D$ has been erased and, according to the UP,
it should be 
impossible to say whether                      
the spin's condition at $t_1$ was $up$ or $down$. 
A reader, worried about the collapse of the  wave function, may find some comfort in noting that
the calculation of probabilities is reduced to evaluation of matrix elements of unitary operators
in the system's Hilbert space, and the collapse problem is not mentioned at all. 

The UP rule does not rely on human experience. Mere existence of a photon, scattered near one of the slits, and carrying a record 
of the electron's past, is enough to destroy the interference pattern, even if the photon is never observed. 
\newline
Feynman's advice was to accept the above rules as the definitive content of the quantum theory, and avoid in this way the \e{blind alley}, 
reserved for those who ask \e{how can it be like that?}\cite{FeynC}. We follow the advice, and ask instead a simpler question,
are the  recently discovered quantum \e{paradoxes} 
mere illustrations of the Feynman's rules?
Some {of them, such as those in [2-5],} certainly appear to be just that. With one of the two holes closed, an electron is able to arrive where it could not arrive 
with both of them open. For this reason, the amplitudes should not be used to prove that the electron was at a particular 
place in the past \cite{FOOT}. This explains the \e{quantum paradoxes} constructed in \cite{3box}-\cite{phot} (for more details see \cite{DSa}).
Would it also be true for the cases discussed in \cite{bruckner}-\cite{Stein}?
\section{The question of consciousness}
Recent work on various versions of the Wigner's friend scenario \cite{bruckner}-\cite{renner} rarely mentions the consciousness of the participants explicitly. 
However, the original Wigner's analysis  \cite{Wig} was motivated by his concern about the role played by consciousness in quantum measurements,
and we must attend to the question before we proceed.
According to Wigner \cite{Wig}, Eq.(\ref{01}) is valid when it describes an inanimate object, but is unacceptable if $|\D(up/down)\ra$
is understood as a state of conscious Wigner's friend, ($F$), who has just seen an outcome $up$ or $down$. 
Would the Feynman's rules of the previous section apply in the presence of intelligent Observers?
\newline
There are at least three ways to look at an Observer's (e.g. $F$', $W$'),  consciousness, while maintaining that quantum mechanics
 should apply to all physical objects, regardless of their size and complexity. 
 \newline
(a) Condition of a consciousness can be deemed to be fully determined by that  of the inanimate \e{physico-chemical
substrate} \cite{Wig} of an Observer's organism. 
Then Wigner's objection can be dismissed by noting that in  Eqs.(\ref{01}) the states $|\D(up)\ra$ and $|\D(down)\ra$ are used for calculating the odds on $F$ seeing $up$ or $down$, 
but never both $up$ and $down$ at the same time. 
However, the difficulty, resolved in this manner, returns when one considers Eqs.(\ref{05})-(\ref{08}). 
Then an outcome $ok$, seen by Wigner ($W$), who performs the second measurement, would leave
the joint system, which now includes the friend $F$,  in an undesirable state of \e{suspended animation} \cite{Wig},
 $\alpha |\D(up)\ra\ot |up\ra+\beta |\D(down)\ra\ot |down\ra$.
\newline
(b) A different problem arises if one were to try placing Observer's \e{extra-observational intellectual inner life} \cite{vN} outside the scope of quantum theory. 
Now, after having seen an $up$, $F$ becomes permanently aware of his/her outcome and this information
resides in a place obscure to quantum reasoning.  
Then if $W$ duly entangles all material records produced by $F$, his probability, $P(ok)$ in Eq.(\ref{08}) should contain an interference term
$2\R[A^{S^*}(ok \gets up\gets  s_0)A^{S}(ok \gets down\gets  s_0)]$.
But if 
$F$ were able to declare that his/her outcome was $up$, the Uncertainty Principle would be violated, since $W$'s interference picture would have to co-exist 
with $F$'s \e{which way?} information. 
\newline
(c) The third possibility \cite{WF-EPL}, \cite{DS_ENT} is a compromise between the first two, 
and is consistent with Feynman's analysis of the double slit experiment.
As in (b), $F$'s consciousness should not be  analysed by means of quantum theory. However, unlike in (b), an Observer is not permanently 
aware of his/her outcome but needs, when necessary, to consult a material record such as the one  kept by his/her memory or in a note, 
accessible to human senses. 
A mere act of perception on the part of an Observer adds nothing to quantum analysis, but a new record  produced as a result
(recall Feynman's photon destroying the double slit interference), needs to be taken into account. As in (a) only material objects need to be subjects of a quantum analysis.  
\newline
Accepting  (c) has further implications.
In particular, at any given time, 
the facts about the outcomes of past experiments are contained in the set of  material records. 
The facts are objective in the sense of being equally valid to all intelligent agents.
 We note that Feynman \cite{FeynL} goes even further in suggesting that facts do not need to be verified, in order to make the presence of records  felt.
 For example, the mere existence of an additional spin, prepared by $F$ in a state $|up\ra$ 
 after seeing an $up$, would remove the interference term from $W$'s probabilities in Eq.(\ref{08}) even if this second spin 
 were to be sent to the Alfa Centauri, and  never seen by anyone again.
 A record can be created, as well as destroyed, in which case the information it contains may be 
irretrievably lost.  
  Observers are free to decide which experiments
 to make, and can sometimes use the arrangements already provided by Nature, as would happen, for example,
with  Wheeler's \e{cosmic interferometer} \cite{Wheel}. 
 Finally, as suggested by Wigner, Observer's consciousnesses \e{never seem to interact with each other directly, but only via  the physical world}\cite{Wig}, 
 i.e. via the world of meaningful records. 
\newline
Throughout the rest of the paper, we will continue to assume that 
quantum theory is valid for all material objects,  and leave the Observers all but outside our discussion 
in accordance with  (c). Next we turn to the extended Wigner's Friend scenario, 
recently proposed in \cite{renner} in order to expose alleged inconsistencies of the 
elementary quantum mechanics.
\section{Extended Wigner's friend scenarios and quantum postmodernism}
With what has just been said, the original Wigner friend scenario fits the description of the double-slit experiment (\ref{01}), 
where the states  $|\D(up/down)\ra$ refer to all material records produced by $F$. One can, however, talk himself into a logical contradiction 
with the help of
the wave function in Eq.(\ref{01}). 
Indeed, measuring the spin, $F$ obtains a definite outcome $up$ with the probability (to shorten the notations, we put $\u^S(t',t)=1$)
$\la \Psi(t_1)|up\ra \la up|\Psi(t_1)\ra=|\la up | s_0\ra|^2$. On the other hand, the probability of $W$'s outcome $ok$ in Eq.(\ref{08}), 
$\la \Psi(t_1)|ok\ra \la ok|\Psi(t_1)\ra$, contains an interference term $2\R[\la s_0|down\ra\la down|ok\ra \la ok|up\ra\la up|s_0\ra]$. This should leave the question \e{$up$ or $down$?} without an answer, yet $F$ appears to know that the answer was $up$.
The reader may be able to see through this apparent {contradiction}. 
We will give a full explanation after considering the case where a similar \e{contradiction} appears in a yet more dramatic form.
\newline
In what has become known as the extended Wigner's friend scenario (EWFS) \cite{renner}, two two-level systems, called  \e{the coin} (C),  and  
\e{the spin} (S), respectively, are prepared in a product state at $t=t_0$. At $t_1$ an Observer $\F$ measures the coin in a basis
$|heads/tails\ra$, and at $\tau> t_1$ the coin is entangled with the spin. At a $t_2> \tau$, $F$ measures the spin in a basis
$|up/down\ra$. The experiment is finished when at $t_3>t_2$ external Observers, $\W$ and $W$, 
measure the entire material content of
 $\F$'s and $F$'s labs using bases
\begin{eqnarray} \label{11}
|\x {Fail}/\x{Ok}\ra = \n
[|\x D(heads)\ra\ot |heads\ra
\pm|\x D(tails)\ra\ot |tails\ra]/\sqrt 2,\n
| {Fail/Ok}\ra = \n
[|D(up)\ra\ot |up\ra
\pm|D(down)\ra\ot |down\ra]/\sqrt 2. 
\end{eqnarray}
\newline 
The initial state and the interaction at $t=\tau$ are chosen so that the state of $\F$'s and $F$'s labs 
just before $\W$ and $W$ make their measurements is given by 
\begin{eqnarray}  \label{12}
|\Phi\ra=
 [|\x D(tails)\ra\ot|tails\ra \ot |D(up){\ra}\ot |up\ra\q\q\q\n
 +|\x D(heads)\ra\ot|heads\ra \ot |D(down){\ra}\ot |down\ra\q\n
+|\x D(tails)\ra\ot|tails\ra \ot |D(down){\ra}\ot |down\ra]/\sqrt{3},
\end{eqnarray}
and can be used to evaluate four different probabilities. 
The likelihood of $\F$ and $F$ seeng the outcomes $heads$ and $up$ is zero, 
\begin{eqnarray}  \label{13}
P(heads, up) =\la \Phi|heads\ra\la heads|\ot |up\ra\la up|\Phi\ra=0.
\end{eqnarray}
A simple calculation shows that also vanish the probabilities 
of $\F$ and $W$ seeing $tails$ and $Ok$, and 
of $F$ and $\W$ seeing $down$ and $\x{Ok}$, 
\begin{eqnarray}  \label{14}
P(tails, Ok) =\la \Phi|tails\ra\la tails|\ot |Ok\ra\la Ok|\Phi\ra=0,
\end{eqnarray}
\begin{eqnarray}  \label{15}
P(down, \x{Ok}) =\la \Phi|down\ra\la down|\ot |\x{Ok}\ra\la\x{Ok}|\Phi\ra=0.
\end{eqnarray}
Equations (\ref{13})-(\ref{15}) suggest that $\W$ and $W$ will never see $\x{Ok}$ and $Ok$, at the same time. 
Yet, a similar calculation shows that the outcomes $\x{Ok}$ and $Ok$ will occur together in about $1/12$ of all trials, 
\begin{eqnarray}  \label{16}
P(\x{Ok},Ok) =\la \Phi|{Ok}\ra\la{Ok}|\ot |\x{Ok}\ra\la\x{Ok}|\Phi\ra=1/12.
\end{eqnarray}
All four probabilities in Eqs.(\ref{13})-(\ref{16}) are legitimate results, and the apparent contradiction needs to be resolved in one way or another.
\newline 
Frauchiger and Renner \cite{renner} were quick to tell their readers that this is where quantum mechanics looses the plot, by making too many conflicting  predictions where only one is required. 
\newline
Alternatively, one may assume that all \e{facts}, as expressed by Eqs.(\ref{13})-(\ref{16}) are indeed valid at the same time, but not to everyone.
In other words, they refer to private perceptions of the participants, held in \e{sealed} laboratories. 
With the labs isolated from each other, there is no danger of comparing the conflicting outcomes, and all perceived results
are equally valid or, if one prefers, equally invalid. The view that quantum theory may only describe such \e{observer-dependent} facts
was proposed by Brukner \cite{bruckner} and found further support, e.g., in \cite{wiseman}.
\newline
There is, however, no need for a radical departure from the standard textbook rules \cite{FeynL}. 
The \e{contradiction}, discussed in the first paragraph of this Section, is a spurious one. The probabilities in Eqs.(\ref{03}) and (\ref{08})
refer to two mutually exclusive scenarios, in which $W$ either erases all records produced by $F$, or preserves them.
Like the proverbial cake, a record cannot be both present and destroyed, and the results (\ref{03}) and (\ref{08}) should never 
be played against each other (we would like to avoid using an over-used  term \e{contextual paradox}). 
The wave function (\ref{01}) just before $W$'s measurement contains no information about course of action $W$ is about to take, 
and contains the answers for each of the $W$'s arrangements. It remains one's own responsibility to decide which
one to use. 
\newline
\begin{figure}
\onefigure[angle=0,width=6.5 cm, height= 4cm]{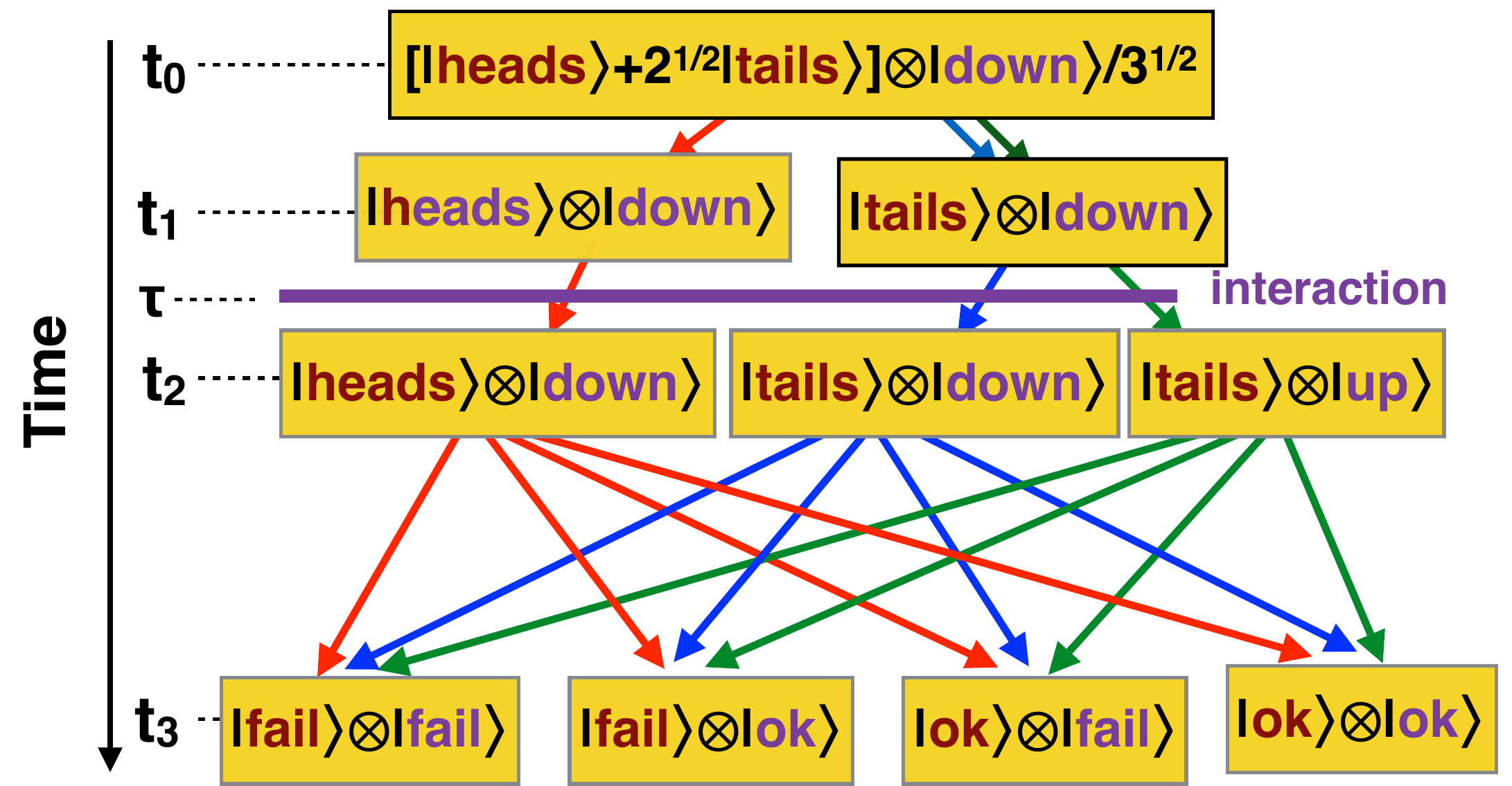}
\caption{Twelve virtual paths of a joint system $\{coin+spin\}$ in extended Wigner's friend scenario.}
\label{fig.1}
\end{figure}
Precisely the same 
happens when  the number of participants is increased to four. 
With the choice of the initial state, and the coupling between the coin and the spin, 
the system maps onto a three-slit setup, where each of the four \e{points on the screen}
can be reached via three virtual paths (see Fig.2 and  Ref. \cite{EWFS} for more details).
Now $\W$'s  and $W$'s may either erase or preserve $\F$'s and $F$ records, respectively. 
There are four exclusive scenarios, and one easily finds \cite{EWFS} that
\newline
i) Eq. (\ref{13}) holds true if both records are preserved, 
\newline
ii) Eq. (\ref{14}) is valid, provided only $\F$'s record survives,
\newline
ii) Eq. (\ref{15}) is valid, provided only $F$'s record survives.
\newline
Finally, Eq.(\ref{16}) applies when both records are erased, and all information about the 
past is irretrievably lost. 
Again, there is no contradiction if one follows Feynman's rules of Ref.\cite{FeynL}.
Next we consider a different case where Feynman's analysis of the double-slit conundrum also 
plays a crucial role. 
\section{The search for the  \e{tunnelling time}}
 The amount of time it takes a {\it classical} particle to cross a given region of space is a useful quantity. 
 In quantum tunnelling, a particle enters and leaves  the barrier region, 
 so it is only natural to assume that it spends there some 
duration $\tau$. 
  In a recent Nature publication \cite{Stein}  
  the authors
put a number 
 on the duration spent by a tunnelling atom in the barrier, and considered the issue \e{resolved}. 
\newline
Well, not quite, as we will show next. 
There are at least two approaches measuring the duration spent by a quantum particle in a given region of space $\Om$.
  One, originally proposed by Baz' \cite{L1}, and recently used in \cite{Stein}, consists of equipping the particle
  with a magnetic moment (spin) which rotates in a small magnetic field, confined to the $\Omega$ . 
  For a classical particle, a ratio of the rotation angle $\varphi$ to the Larmor frequency $\omega_L$ yields the desired result 
 $\tau_{cl}=\varphi/\om_L$.
 \newline
 For a quantum particle, making transition between initial and final states $|\psi\ra$ and $|\phi\ra$ over a time $T$, one finds many possible durations, 
 and  is able to define the corresponding probability amplitudes $A(\phi\gets \tau \gets \psi)$ \cite{L3} 
 by summing the amplitudes $\exp\{iS[x(t)]\}$ \cite{FeynH} over the Feynman paths, spending in $\Om$ precisely $\tau$ seconds. 
With this, a non-relativistic  transition amplitude can be seen to result from interference between all allowed durations
 \begin{eqnarray}  \label{21}
\la \phi|\u(T)|\psi\ra=\int_0^T A(\phi \gets \tau \gets \psi)d\tau.
\end{eqnarray}
With many virtual durations in place, the final state of the spin, travelling with the particle,  will be a superposition $\int_0^T |\varphi=\om_L\tau\ra A( \tau \gets \psi)d\tau$
where $|\varphi\ra$ stands for the initial spin's state, rotated by an angle $\varphi$ around the direction of the field.
Notably, a {\it sum}  of such rotations is not a new rotation
\textcolor{black}{(}there is no single angle to divide by $\om_L$\textcolor{black}{)},
and the Larmor clock method meets with its main difficulty (see Fig.3b).
\begin{figure}
\onefigure[angle=0,width=7.0 cm, height= 5.5cm]{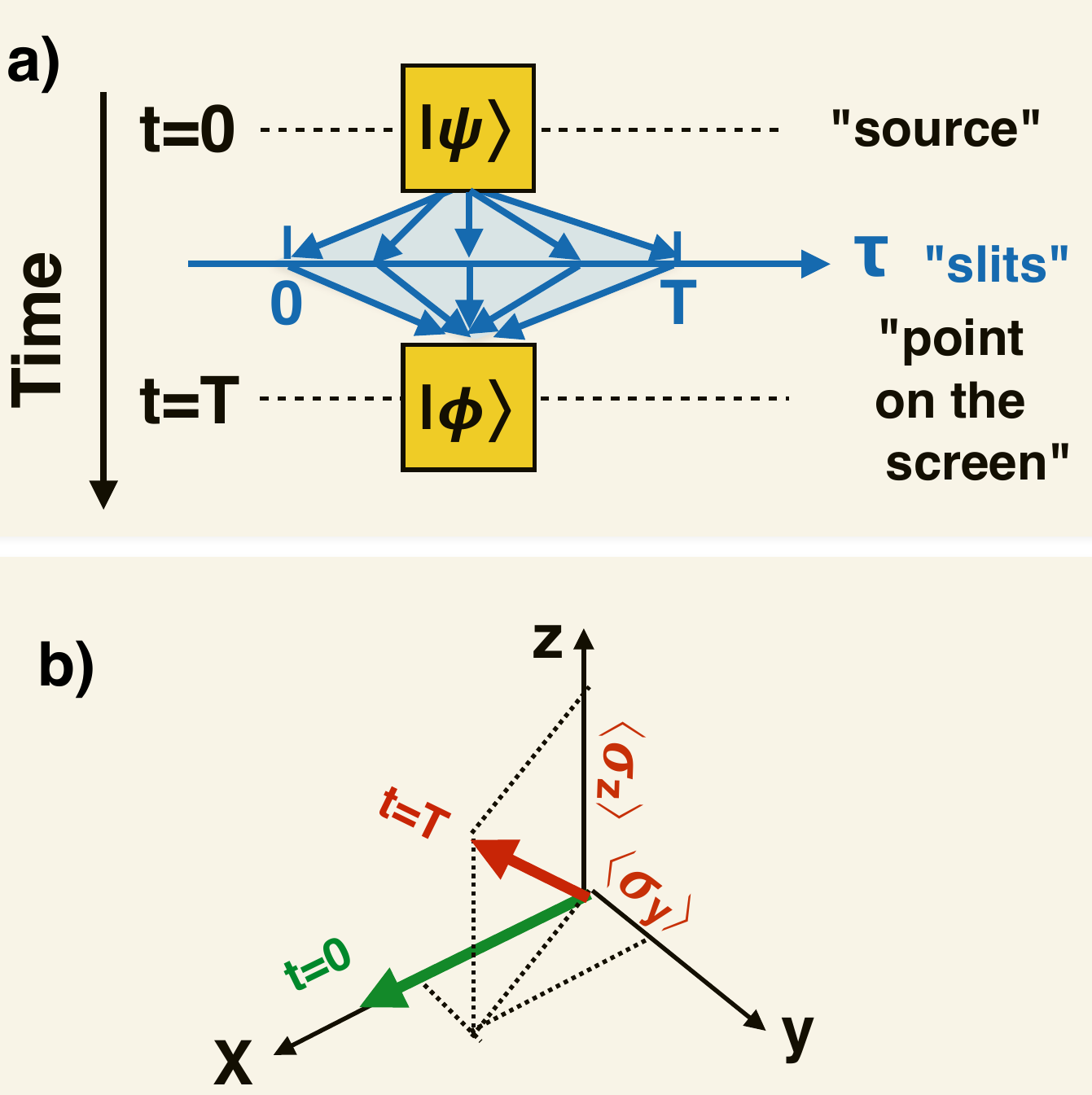}
\caption{a) A non-relativistic quantum particle can reach its final state $|\phi\ra$ by spending 
any virtual duration $0\le \tau \le T$ in a specified region of space $\Omega$. 
b) In each virtual scenario, a spin-$1/2$ initially polarised along the $x$-axis rotates in the $xy$-plane 
by $\varphi =\tau \omega_L$. 
However, a superposition of such rotations is not a new rotation in the $xy$-plane,
 and at $t=T$ the spin acquires also a non-zero $z$-component.
 For a weakly perturbing Larmor clock one finds 
 $\la \sigma_y(T)\ra \approx \R[\tau_w]$ and  $\la \sigma_z(T)\ra \approx \Ip[\tau_w]$, 
 with $\tau_w$ defined in Eq.(\ref{37}).
 }
\label{fig.1}
\end{figure}
\newline
Equation (\ref{21}) describes a multi-slit problem where a particle can reach the \e{point on the screen}, $|\phi\ra$, 
by passing through a continuum of \e{slits}, labelled by $\tau$, as shown schematically in Fig.3a.
As in the original double-slit case, one has a choice between keeping the transition intact, and not knowing the duration $\tau$, 
or measuring $\tau$ at the cost of destroying the interference, and with it the studied transition \cite{L2}-\cite{L4}. 
The only exception is the classical case, where $A(\phi\gets \tau \gets \psi)$
rapidly oscillates everywhere but in the vicinity of the classical value  $\tau=\tau_{cl}$, 
and $\om_L\tau_{cl}$ defines a unique angle by which the spin rotates.
In a classically forbidden tunnelling transition
 no unique duration can be selected. To make the matter worse, the $A(\phi\gets \tau \gets \psi)$
 is not itself small, and the very small tunnelling amplitude in the l.h.s. of Eq.(\ref{21}) is a result of a
 very precise cancellation. Any attempt to perturb this delicate balance is, therefore, likely to destroy tunnelling.
 \newline
The second method to measure the delay, experienced by a particle in a potential, is to compare the particles's final position 
with that of its free moving counterpart \cite{McColl} (see Fig.4). 
\begin{figure}
\onefigure[angle=0,width=6.0 cm, height= 2.5cm]{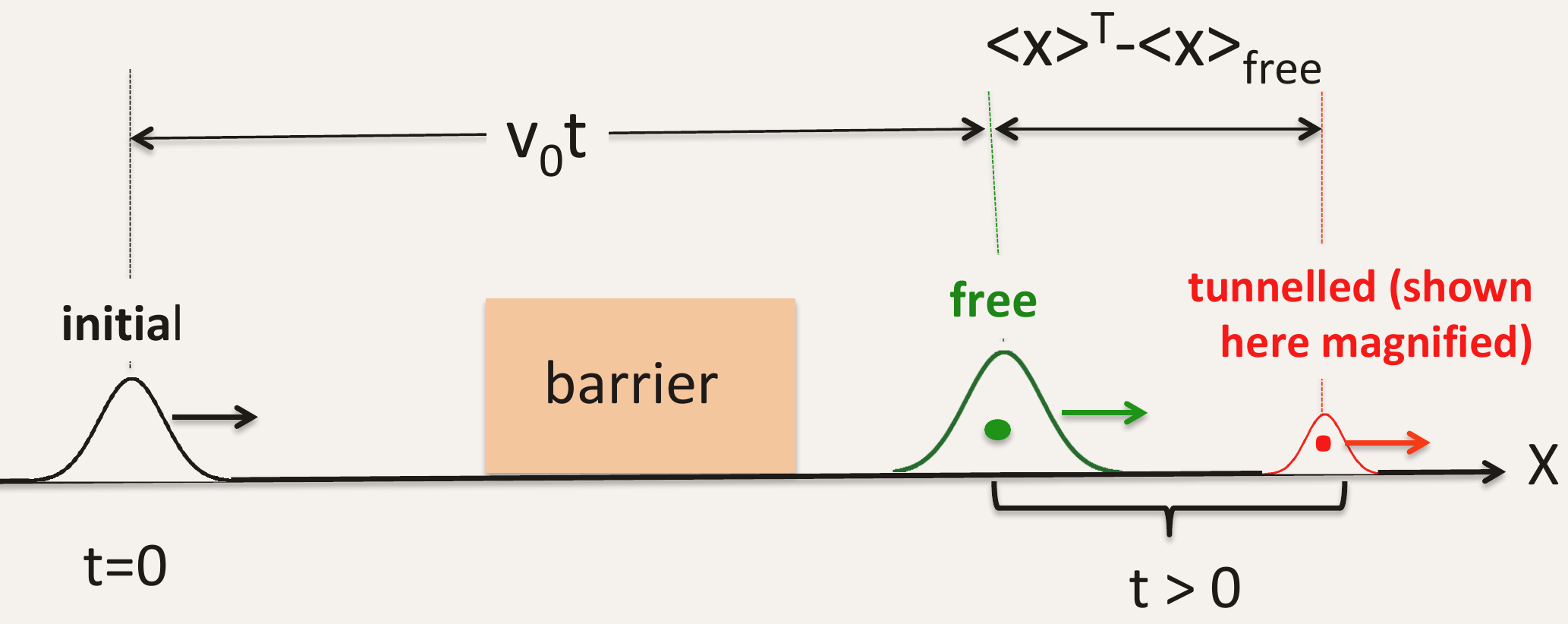}
\caption{The distance between the centre of mass  of a tunnelling wave packet 
and that of its freely propagating counterpart. An attempt to deduce from it the duration spent in 
the barrier meets with the same  difficulty as the application of a Larmor clock shown in Fig.3b.} 
\label{fig.1}
\end{figure}
For a classically forbidden transition, it meets with the same difficulty. 
The state of a particle with a momentum $p$ and energy $E$, transmitted across a finite range barrier, or  well, is given by $T(p)\exp(ipx)$, 
where $T(p)$ is the
barrier's  transmission amplitude. Using a Fourier transform $\xi (x') =(2\pi)^{-1}\int_{-\infty}^\infty T(p')\exp(ip'x')dp'$,
one can write the transmitted state  as a superposition of plane waves, each displaced in space by a distance $x'$ \cite{WP1}, \cite{WP2}
\begin{eqnarray}  \label{22}
T(p)\exp(ipx)=\int dx' \xi(x')\exp[ip(x-x')].
\end{eqnarray}
A similar expression exists  for a transmitted wave packet with a mean momentum $p_0$, 
\begin{eqnarray}  \label{23}
\psi^T(x,t)=\exp[ip_0x-iE(p_0)t]
\times\n
\int G_0(x-x',t)\eta(x',p_0)dx',
\end{eqnarray}
where $\eta(x',p_0) \equiv \exp(-ip_0x') \xi(x')$, and $G_0(x,t)$ is the envelope of the same initial wave packet, as it would be {in the absence of the potential}, 
\begin{eqnarray}  \label{24}
\psi_0(x,t)\equiv \exp[ip_0x-iE(p_0)t] G_0(x,t).
\end{eqnarray}
In general, a sum of spatial shifts is not a new shift, and no single shape is selected from the collection
of the envelopes $G(x-x',t)$ in Eq.(\ref{23}).
Again, the only exception is  the classical limit, where an oscillatory $\eta(x',p_0)$ 
develops a stationary region around  a classical value $x'=x'_{cl}(p_0)$, 
and one recovers  the classical result $\psi^T(x,t)\approx \psi_0(x-x'_{cl},t)$. 
For a barrier not supporting bound states, $\xi (x')\equiv 0$ for $x'>0$ \cite{WP1}, so that none of the envelopes 
 in (\ref{23}) are {\it advanced}, relative to the freely propagating $G_0(x,t)$.
 In a tunnelling regime, 
 the spatial delay 
 experienced  by the particle is lost to interference  \cite{WP2}, just like the duration $\tau$ of the previous example.
 The reader may reasonably ask if,
 according to the Uncertainty Principle, 
 a  tunnelling time cannot exist, how could McColl conclude \cite{McColl}
  that tunnelling is a delay-free process?
 And what was measured in the experiment reported in \cite{Stein}? 
\section{Complex times and weak "measurements"}
To answer these questions, we consider again the double-slit case
(\ref{01})-(\ref{04}), but this time with the probe $\D$ replaced by a von Neumann pointer \cite{vN}
with position $x$, 
prepared in a state $|G_0\ra$, $\la x |G_0\ra\sim \exp(-x^2/\Delta x^2)$.  The pointer 
is set up to measure an operator $\hat B=|up\ra B_1 \la up| + |down\ra B_2 \la down|$ with
eigenvalues $B_1$ and $B_2$.
Now the state of the pointer, provided the spin is found in a state $|ok\ra$,
is given by 
\begin{eqnarray}  \label{31}
G(x,t_2)= G_0(x-B_1)A_1
+G_0(x-B_2)A_2,
\end{eqnarray}
where we use a shorthand  $A_1$ and $A_2$ for $A^S(ok\gets up \gets s_0)$ and 
$A^S(ok\gets down \gets s_0)$, respectively.
The probability for the observed system to arrive in $|ok\ra$ is
\begin{eqnarray}  \label{32}
P(ok)=\int |G(x,t_2)|^2 dx =|A_1|^2+
|A_2|^2\n
 +
2\R[A_2^*A_1)]\times\int G(x-B_1)G(x-B_2)dx.
\end{eqnarray}
The pointer carries a record of the spin's condition at $t_1$, and in the {accurate}  limit $\Delta x \to 0$
it is always possible to find out  whether it was $up$ or $down$. In the opposite 
limit, $\Delta x \to \infty$, a pointer's reading $x$ cannot be used to distinguish  
between the paths, and $P(ok) =|A_1+A_2|^2$. The possible pointer 
readings are distributed between $-\infty$ and $\infty$,  and the value of $\hat B$,
 measured in this way, is well and truly {\it indeterminate}. 
\newline 
One can also use Eq.(\ref{32}) to calculate the average pointer reading $\la x\ra$, 
conditional on the system arriving at $|ok\ra$.
For $\Delta x \to \infty$ one obtains \cite{UP}
\begin{eqnarray}  \label{34}
\la x\ra\xrightarrow[\Delta x \to \infty]{}  \R\left[\frac{B_1A_1+B_2A_2}{A_1+A_2}\right]. 
\end{eqnarray}
At first glance, Eq.(\ref{34}) appears to contradict the Uncertainty Principle, 
because a definite value, $\la x\ra$, has been obtained in a situation where
everything was meant to be indeterminate. This is, however, not so, since 
the initial and final states $|s_0\ra$ and $|ok\ra$ can always be chosen 
so as to give the r.h.s. of Eq.(\ref{34}) {\it any} desired value - large, small, positive, 
negative, or zero \cite{UP}. The Uncertainty Principle 
still applies, albeit at a different level. 
Clearly, one  thing that can be learnt from Eq.(\ref{34}) is that  adding the system's  amplitudes, 
multiplied by  $B_1$ and $B_2$, dividing the result by their sum, and taking the real part of the fraction,
gives a particular  number. 
It is far less clear, since Feynman's rules \cite{FeynL} give no clue in this regard, whether Eq.(\ref{34})
can have any \textcolor{black} {other} significance. 
\newline
The non-perturbing \e{weak} limit $\Delta x \to \infty$ was first studied in \cite{AV}, where 
the quantity in brackets in Eq.(\ref{34}) was called the \e{weak value of an operator $\hat B$}, 
equally written as
\begin{eqnarray}  \label{35}
B_w\equiv \frac{\la ok|\hat B|s_0\ra}{\la ok|s_0\ra}. 
\end{eqnarray}
The authors of \cite{AV} made two claims which, while no doubt helping subsequent popularity of the subject \cite{WMrev}, have 
led to a fair amount of confusion, including that surrounding the tunnelling time problem discussed here.
Firstly, it was claimed that Eq.(\ref{35}) \e{defined a new kind of quantum variable}, whereas, as we have seen, 
they were describing a particular combination of the familiar probability amplitudes.
Secondly, they found it surprising that a weak value of a spin-$1/2$ component could take a value of $100$. 
According to the Uncertainty Principle, it would be surprising if it could not.
\newline
It is easy to see what all this means \textcolor{black}{for} the quest to find and measure  \e{the tunnelling time}.
Neglecting the spreading of the wave packet, $G_0(x,t)\approx G(x-v_0t)$, and comparing Eq.(\ref{23}) with Eq.(\ref{31}),
we note that
we are dealing with an inaccurate measurement of the quantity which we earlier described as 
the spatial delay, or shift, $x'$, with which the transmitted particle with a momentum $p_0$ leaves  the barrier.
The particle's own position $x$ plays the role of the pointer, 
and by sending $\Delta x \to \infty$ we can make 
the measurement weak. In this limit, the distance between the centre of mass of the transmitted wave packet
and its freely propagating counterpart is given by an analogue of Eq.(\ref{34}) \cite{WP2}
\begin{eqnarray}  \label{36}
\la x\ra^T - \la x\ra_{free}\xrightarrow 
[\Delta x \to \infty]{}
\R[x'_w]\equiv \q\q\q \n
\R\left [\frac{\int x'\eta(x',p_0)dx'}{\int \eta(x',p_0)dx'}\right] =-\partial_p\Phi(p_0),
\end{eqnarray}
where $\Phi(p)$ is the phase of the transmission amplitude, $T(p)=|T(p)|\exp[i\Phi(p)]$. 
This can be verified experimentally \cite{Nat}, but what can be learnt from such an experiment?
The answer is the same as before: a barrier can be  characterised by distribution of virtual shifts it imposes upon the transmitted particle.  
The measured distance (\ref{36}) is a weighted sum of the corresponding amplitudes. 
\newline
An attempt to give this quantity a deeper  meaning immediately meets with difficulties. For a broad rectangular barrier of a height $V$, width $d$, 
and a non-relativist particle of a mass $\mu$, one finds $T(p,V)\sim \exp(-ipd)\exp(-d\sqrt {2\mu V-p^2})$.
According to Eq.(\ref{36}), the (small) tunnelled pulse is {\it advanced} by roughly the barrier's width $d$, 
no matter how broad the barrier is. This Hartmann effect \cite{Hart} has more dramatic consequences if one tries to deduce 
from Eq.(\ref{36}) the \e{time $\T$ the particle has spent in the barrier.} The result $\T \approx 0$ seems to point towards 
a conflict between quantum mechanics and special relativity \cite{Nimtz}. The problem, however, has no relativistic 
implications. In Eq. (\ref{23}) the envelopes $G(x-x',t)$, from whose front tails the transmitted wave packet is built, are all delayed even relative to 
free propagation, and certainly so relative to the motion at the speed of light.
\newline
Time measurements by means of a weakly perturbing Larmor clock  \cite{Stein},\cite{L1} suffer from the same deficiency \cite{L2} - \cite{L4} .
They inevitably involve a \e{complex time}, ${\tau_w}$ \cite{L4},
\begin{eqnarray}  \label{37}
\tau_w
\equiv \frac{ \int_0^T \tau A(\phi \gets \tau \gets \psi)d\tau}{\int_0^T  A(\phi \gets \tau \gets \psi)d\tau}=\R[\tau_w]+i\Ip[\tau_w],
\end{eqnarray}
 another \e{mean value}, calculated with an alternating
complex value {distribution}, 
which cannot be interpreted as a meaningful duration spent in a given region of space. Indeed, by choosing $|\phi\ra$ so that $\la\phi|\u(T)|\psi\ra \to 0$,  one can always ensure that  $\R[{\tau_w}]$, $\Ip[\tau_w]$ , and $|\tau_w|$ exceed the duration of motion, $T$, 
which rather proves the point. 
A measured value of $\R[\tau_w]$ represents a relation between 
the amplitudes $A(\phi \gets \tau \gets \psi)$ in Eq.(\ref{21}),  while the Uncertainty Principle ensures that the 
 duration
spent in the region where the spin's rotates, remains indeterminate, just like the slit chosen in the double-slit experiment.
We are back to Feynman's  \e{only mystery quantum mechanics}. 
\section{Conclusions and discussion}
In summary, one finds Feynman Uncertainty Principle at the centre of many recent developments in elementary 
quantum mechanics. The principle, we recall, implies that in a an interference phenomenon the system's past 
remains indeterminate, and an inquiry inevitably destroys the phenomenon.  
In \cite{FeynC} Feynman's wrote {\it  \e{They can give you a wider class of experiments than just the two slit interference 
experiment. But that is just repeating the same thing to drive it in.}} Well, they did, 
and often  without mentioning the \e{thing} in question.
Today  \e{Feynman's blind alley} \cite{FeynC} comes adorned with many extraordinary, yet not particularly useful claims. 
Most of them arise  either from mixing incompatible scenarios 
(we promised not to use the words \e{contextual paradox}),
or from giving probability amplitudes a meaning they were not suppose to have in the first place \cite{FOOT}.
\newline
Thus, accepting the general principles of quantum mechanics, as  given in \cite{FeynL}, one notes that the seeming {conflict} between the statements (\ref{13})-(\ref{16}) does not prove that quantum theory is \e{inconsistent} 
but only 
that an interference pattern cannot co-exist with the knowledge of the slit used by an electron.
Neither does it imply that each Observer is entitled to his or her own facts. The probabilities (\ref{13})-(\ref{16}) refer to mutually exclusive situations, when some of the virtual scenarios can be distinguished, 
and some others cannot. 
\newline
The so-called \e{weak measurements}, used in \cite{3box} -\cite{cat},\cite{AV}, quantify particular relations between Feynman's  amplitudes \cite{UP}, but provide no insight
into the amplitudes'  meaning or physical significance beyond what was said in \cite{FeynL}. 
A quantum particle is not in two places at the same time - it is either in one place, or it is impossible to say where it is \cite{FOOT}.
Similarly, using the \e{weak values} as  the evidence of the particle's presence does not prove that photons can be found in a place they never entered \cite{phot}, 
but only that two non-zero amplitudes may sum up to zero \cite{phot1}.
\newline
For the same reason, tunnelling which results from interference between different delays, whichever way one looks at it, 
cannot be described by a single meaningful  duration, and poses no threat to special relativity. Finding such a delay would invalidate the UP, and uproot quantum 
mechanics \e{protected} by the principle \cite{FeynL}. 
This is why the claim 
 to have measured the traversal 
time and {\it  \e{resolved the controversy regarding how long a tunnelling particle spends in the barrier region}} made in \cite{Stein} by Ramos {\it et al} is grossly misleading, even if the experiment itself is technically perfect.
\newline
In summary, it may not be a bad idea to check whether a proposed experiment, 
no matter how spectacular, could be just another illustration of Feynman's 
\e{only mystery} \cite{FeynL}.
If it is
quantum mechanics could only be accused of being quantum mechanics as we know it.
If it is not, something new may be learnt as a result.
Yet by ignoring the Uncertainty Principle altogether one risks arriving 
at a conclusion, which is either wrong, or after scrutiny, will prove to be 
the \e{repetition of the same thing} \cite{FeynC}, in other words, a platitude. 
\acknowledgements
Financial support of
MCIU, through the grant
PGC2018-101355-B-100(MCIU/AEI/FEDER,UE)  and the Basque Government Grant No IT986-16,
is acknowledged by DS.

\end{document}